\documentclass[twocolumn,showpacs,preprintnumbers,amsmath,amssymb,pra]{revtex4}
\usepackage{graphicx}
\usepackage{dcolumn}
\usepackage{hyperref}

\begin{document}
\title{Spontaneously induced emitter-radiation entanglement due to confinement to photonic band gap}

\author{Sintayehu Tesfa} \email{sint\_tesfa@yahoo.com}
\affiliation{Physics Department, Addis Ababa University, P. O. Box 1176, Addis Ababa, Ethiopia}

\date{\today}

\begin{abstract}The study of spontaneously induced nonclassicality as a result of the interaction of an ensemble of two-level emitters embedded onto crystal structure embodying photonic band gap (PBG) is presented. The method of coherent-state propagator is applied upon expressing collective atomic operators in terms of boson operators in view of Schwinger's representation. The autocorrelation of photon number of the radiation alone is shown to display super-Poissonian statistics and chaotic photon character. However, the state of the coupled system is found to exhibit entanglement and nonclassical intensity correlation attributed to the confinement, where intensity of the emitted radiation and degree of entanglement are enhanced near the edge. The prospect of embedding emitters onto PBG can hence be taken as a reliable testing ground for examining fundamentals of emitter-radiation interaction and implementing quantum information processing task that requires mapping of stationary memory with flying message.\end{abstract}

 \maketitle


Research on interaction of electromagnetic field with atom(s) has been one of the central themes in the field of quantum optics \cite{booksc, jmo502765, sariv} and photonics \cite{pra71023812}, where  the corresponding excitation is expected to manifest nonclassical features \cite{josab41490, epjp137301}. For instance, spontaneously induced quadrature entanglement of the combined state of an ensemble of two-level atoms and radiation trapped in the cavity has been reported \cite{jmo551683, oc227349}. It then appears imperative exploring whether confining the electromagnetic field in crystal structure can also lead to emitter-radiation nonclassical correlation? Notably, for a periodic arrays of non-dispersive high dielectric constant medium, the interplay between the energy and momentum of the electromagnetic field can be altered by creating a gap in the density of states over a narrow range of frequency \cite{pra584168, prl582486} that hinders some  range of frequency to linearly propagate in the crystal \cite{josab10283}. The electromagnetic field whose frequency falls in the created gap would then be confined in this structure \cite{prl642418}, which enables the radiation to resonantly coupled to the atoms or impurities \cite{oe141658}. PBG as a result is expected to offer a unique way to tailor propagation of electromagnetic waves; and consequently, render opportunity for controlling and manipulating the generated radiation \cite{np1449}. 

Application of such mechanism has led to several exciting phenomena such as suppression of spontaneous emission \cite{prl582059}, formation of strongly localized state of light \cite{prl582486}, and creation of photon-atom bounded state \cite{prl642418,prb4312772}. Emission-absorption interplay of an ensemble of atoms embedded onto crystal structure embodying PBG has also been shown to significantly shifted from free space as well as corresponding cavity situations \cite{prl795238}. Since the emitted radiation can be confined and behaved in the same way as in cavity setup, a strong coupling that can be established via successive emission-absorption exchange is envisaged to lead to significant cooperative phenomenon and emitter-radiation entanglement \cite{pla3733413, epjp137301}. Taking this as motivation, the ensuing correlation between the state of an ensemble of two-level emitters embedded onto one-dimensional array,  as impurity or hole or defect via changing the atomic makeup of the dielectric material \cite{prl79205}, and electromagnetic field trapped in PBG would be explored. Coherent emission from similar scheme for instance has been analyzed near the edge in non-Markovian regime in terms of Langevin equation, where the characteristics of the population is found to be radically different from cavity scenario \cite{pra584168, prl743419}. Even then, assuming the noise fluctuations to make temporal distinctions difficult, we opt to resort to Born-Markov approximation and stick to Dicke's model with small sample limit \cite{books}. The emitters are then denoted by spin operators that can be mapped to boson operators by using Schwinger's representation of the angular momentum \cite{pra423051}.

Pertaining to the assessment that matter-light entanglement ensures storage and retrieval of single photons between remote quantum memories that paves the way for distribution of quantum information, the anticipated modification in the emission-absorption interplay and recent proposal of using two-level emitters embedded onto PBG for quantum computing \cite{jap865237} is taken as foundation to pursue this study. We as a result seek to explore wether the absorption of radiation emitted by ensemble of two-level emitters followed by spontaneous emission leads to nonclassical correlation. To this effect, once the collective spin operators are mapped onto boson operators \cite{prl96030404}, the $Q$-function is derived in interaction picture following the approach outlined in \cite{books, pra465379}. With the help of obtained quasi-statistical distribution, the degree of induced entanglement is quantified in view of the approach introduced in \cite{pra74043816} and photon statistics of the radiation are sought for. Since physical mechanisms other than photon exchange  are not taken into consideration, the emerging entanglement is linked to the process of spontaneous emission \cite{jmo551683}.
 
\begin{figure}[htb]
\centerline{\includegraphics [height=6.5cm,angle=0]{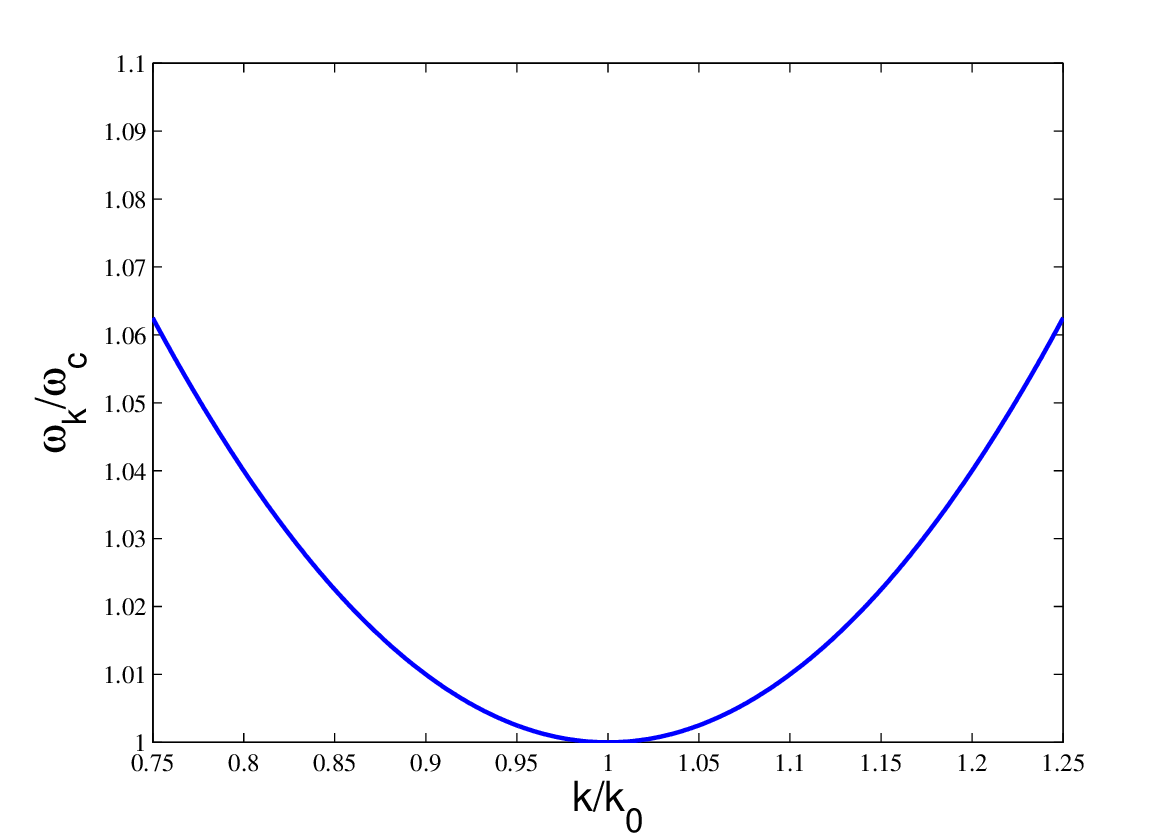}}
\caption {\label{fig1}Dispersion relation.} \end{figure}

Even though resonant frequency is presumed to be very close to edge frequency \cite{pra584168, job6715}, the density of modes can vary with frequency in a manner consistent with dispersion relation. Such a change is anticipated to affect the nature of emitter-radiation interaction even to the extent of forging nonclassical correlation. To explore to what extent this dependence modifies the characteristics of the dynamics, effective mass approximation is evoked in which the edge is designated by ${\vec{k}}={\vec{k}}_{0}$, where ${\vec{k}}_{0}$ is the radius of the sphere in ${\vec{k}}$ space about which the expansion is performed \cite{prl743419}. The frequency can then be expanded for isotropic model near the edge \cite{pra64033801} that leads in a one-dimensional case \cite{prl743419, pra584168} to 
\begin{align}\label{drel}\omega_{k}=\omega_{c}\left[1+\left({k\over k_{0}}-1\right)^{2}\right],\end{align}
where $\omega_{c}$ is the frequency at the edge \cite{prb4312772, pra501764, prl79205, pra423051}. It is possible to see  near the edge ($0.75k_{0}\le k\le1.25k_{0}$), the frequency of the confined radiation is within 6\% of the higher or lower edge frequency \cite{prl631950} (see also Fig. \ref{fig1}). It is also acceptable to adjust the range of the frequency well above this value via changing the makeup of the crystal structure or selecting appropriate atomic transition frequency based on which edge one seeks to approach. Even then, it is worth noting that the value of $k/k_{0}$ is set to be consistent with the expansion made near the edge while deriving the dispersion relation \cite{np1348}.

The two-level emitters on the other hand can be characterized by excitation frequency $\omega_{a}=E_{a}-E_{b}$, where $E_{a}$ and $E_{b}$ denote energy eigenvalues of the upper ($|a\rangle$) and lower energy ($|b\rangle$) states that can be designated as electric dipole. Emitter-radiation coupled system can as a result be considered as interaction of electric dipole moment with confined electromagnetic field. Specifically, point interaction approximation is envisioned in which spatial extent of the active region is taken to be less than the wavelength so that dipole-dipole interaction near the edge is neglected. Density of emitters is also assumed to be very small to neglect direct emitter-emitter interactions. Such simplified version is chosen with intention of grasping quantitative picture of near band edge phenomena rather than exploring detailed population dynamics \cite{pra584168}.

The Hamiltonian that describes a collection of $N$ two-level emitters embedded onto crystal structure; and subsequently, interact with confined single-mode radiation can be epitomized in the rotating wave approximation by \begin{align}\label{ham}\hat{H}_{I}=ig(\omega_{k})\big(\hat{J}_{12}\hat{a}
-\hat{a}^{\dagger}\hat{J}_{21}\big),\end{align} 
where
\begin{align}\hat{J}_{12}=\sum_{j=1}^{N}|a_{j}\rangle\langle b_{j}|;~~~~\hat{J}_{21}=\sum_{j=1}^{N}|b_{j}\rangle\langle a_{j}|\end{align}
are collective spin operators, $\hat{a}$ and $\hat{a}^{\dagger}$ are annihilation and creation operators \cite{jap865237, prb4312772, prl743419, pra584168}. $g(\omega_{k})$ on the other hand stands for coupling of quantum states of the emitters with radiation and taken to have the form
\begin{align}g(\omega_{k})=\omega_{a}\mu_{ab}\sqrt{{1\over2\varepsilon_{0}V\omega_{k}}},\end{align} 
where $\mu_{ab}$ is pertinent dipole moment, $\varepsilon_{0}$ is permittivity of free space, $V$ is quantization volume, and $\omega_{k}$ is frequency of the radiation that depends on dispersion relation.

As one can see from Eq. \eqref{ham},  since spin and boson operators are governed by different algebras, studying the corresponding correlation is not always straightforward. In light of this, collective spin operators are expressed  in terms of boson operators in view of Schwinger's representation of angular momentum operators as
\begin{align}\label{ape1}\hat{J}_{12}=\hat{b}^{\dagger}\hat{c};~~~~ \hat{J}_{21}=\hat{c}^{\dagger}\hat{b},\end{align}
where $\hat{b}$ and $\hat{c}$ are related to atomic states of the lower and upper energy levels \cite{booksc, pra423051, pra501764}. Interaction of ensemble of two-level emitters with single-mode radiation can then be designated by
\begin{align}\label{ape3}\hat{H}_{I}=ig(\omega_{k})\big(\hat{a}^{\dagger}\hat{b}^{\dagger}\hat{c} -\hat{a}\hat{b}\hat{c}^{\dagger}\big).\end{align}
The manner in which the operators appear in Eq. \eqref{ape3} unravels that annihilation of quantum state in the upper energy level leads to simultaneous creation of a photon and single occupancy of the lower energy level.

To overcome the difficulty of solving coupled differential equations following from trilinear Hamiltonian, the case when almost all the emitters are initially excited is considered. It is hence common knowledge treating the operator $\hat{c}$ as a $c$-number $\gamma$ set to be real-positive constant \cite{pra211297}. With this background, Eq. \eqref{ape3} reduces to
\begin{align}\label{apeh}\hat{H}_{I}=iG(\omega_{k})\big(\hat{a}^{\dagger}\hat{b}^{\dagger} -\hat{a}\hat{b}\big),\end{align}
where $G(\omega_{k})=g(\omega_{k})\gamma$ accounts for the strength of pumping. It is then admissible to analyze quantum features and statistical properties of the radiation by solving the associated differential equations following from Heisenberg's equation or applying the corresponding quasi-statistical distribution. In connection to the fact that equations resulting from quasi-statistical distribution can be handled in view of classical formulation and 
following the procedure outlined in \cite{booksc, pra465379}, the $Q$-function that corresponds to Hamiltonian \eqref{apeh} is found to have the form
\begin{align}\label{ape4}Q(\alpha,\beta,t)&={1\over\pi^{2}\cosh^{2}(G(\omega_{k})t)} \exp\big[-\alpha^{*}\alpha-\beta^{*}\beta \notag\\&+\big(\alpha^{*}\beta^{*}+\alpha\beta\big)\tanh(G(\omega_{k})t)\big],\end{align}
where $\alpha$ and $\beta$ are $c$-number variables associated with radiation and occupancy of the lower energy level \cite{jmo551683}. It may not be difficult to notice  that $Q$-function \eqref{ape4} resembles that of two-mode parametric optical oscillator except that the coupling depends significantly on dispersion relation.


Once the $Q$-function is obtained, it is straightforward to find the intensity of the accompanying emitted radiation that can be defined as
\begin{align}\hat{n}_{r}(t)=\hat{a}^{\dagger}(t)\hat{a}(t),\end{align}
where the corresponding mean photon number turns out using Eq. \eqref{ape4} to be
\begin{align}\label{meanr} \langle\hat{n}_{r}(t)\rangle=\sinh^{2}(G(\omega_{k})t).\end{align}
The dynamics of the occupancy of the lower energy level can also be derived in the same way. In view of the symmetry of operators in Eq. \eqref{apeh}, mean number of emitters that occupy the lower energy level would be
\begin{align}\label{meana} \langle\hat{n}_{a}(t)\rangle= \langle\hat{b}^{\dagger}(t)\hat{b}(t)\rangle=\sinh^{2}(G(\omega_{k})t).\end{align}

\begin{figure}[htb]
\centerline{\includegraphics [height=6.5cm,angle=0]{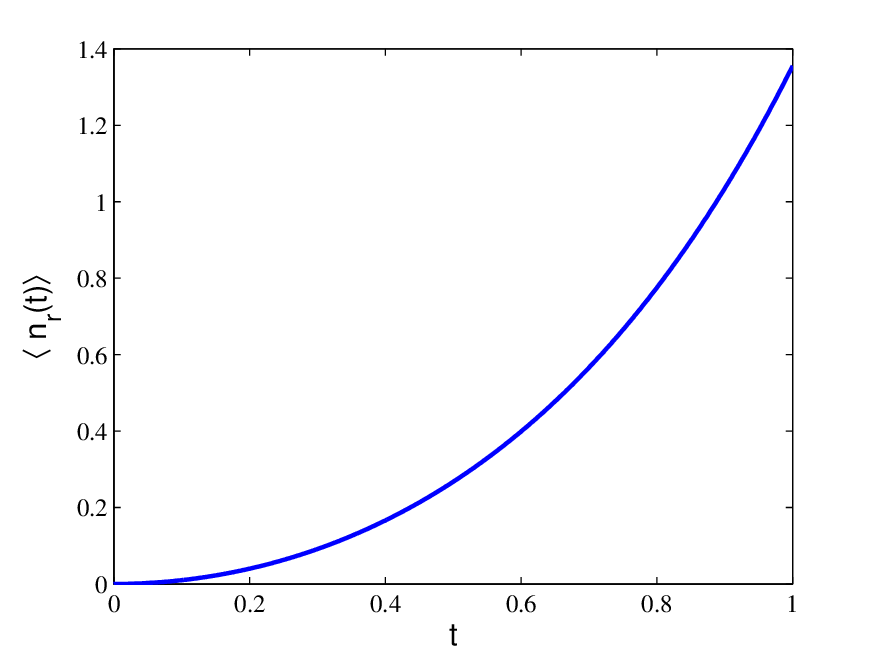}}
\caption {\label{fig2}Mean number of the radiation emitted to photonic band gap against time for $k = 1.12k_{0}$ to be consistent with expansion made near the edge while deriving the dispersion relation.} \end{figure}

This outcome ensures that the mean photon number of the radiation emitted to PBG is equal to the number of emitters de-excited to the lower energy level. This phenomenon is the reminiscence of the fact that annihilation of a single upper energy state gives rise to creation of one photon and occupancy of a single lower energy state, which holds true when there is no damping and/or dephasing. Since more emitters prefer to participate in the emission process with time, it would be judicious expecting enhanced intensity of light at steady state as shown in Fig. \ref{fig2}. This observation entails that such a scheme can be applicable to amplify light in case one manages to externally pump, which can be taken as motivation for pondering on such arrangement while seeking to generate intense radiation in miniaturized system as in lasing mechanism \cite{pra69013816}.

\begin{figure}[htb]
\centerline{\includegraphics [height=6.5cm,angle=0]{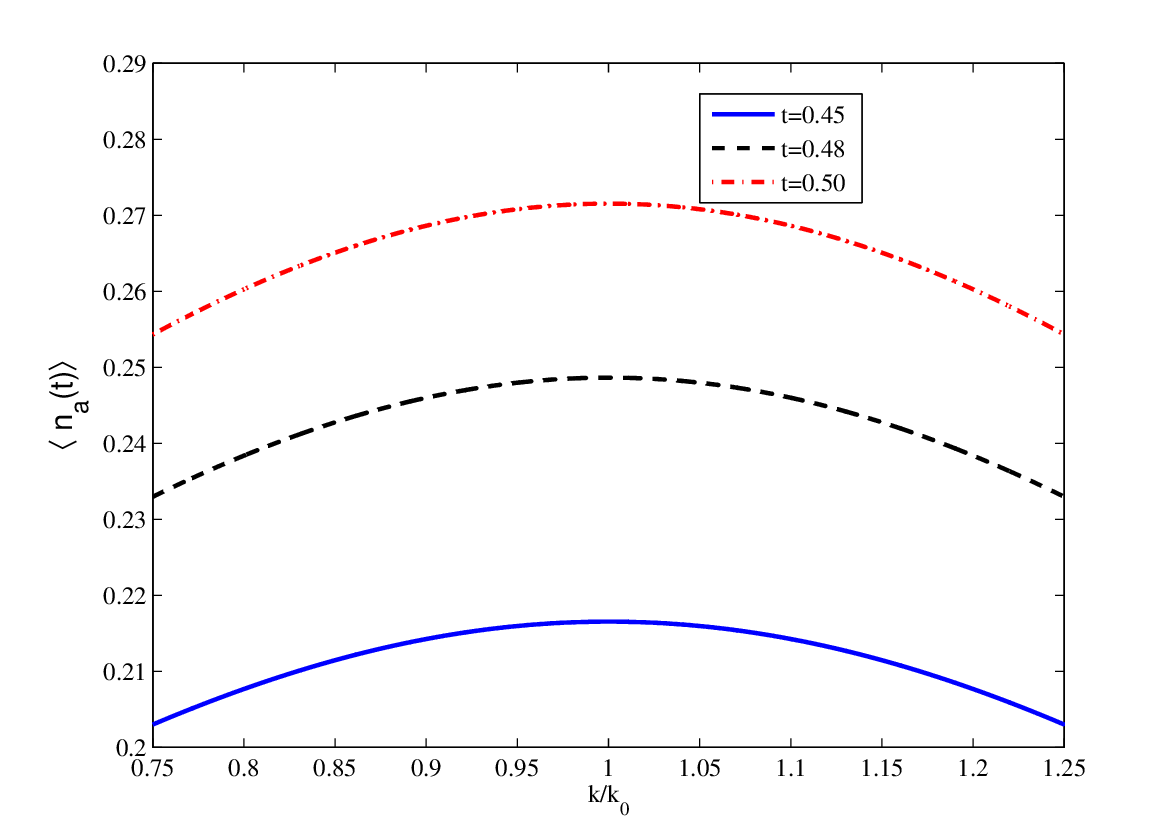}}
\caption {\label{fig3}Mean number of emitters de-excited to the lower energy level against $k/k_{0}$ at different times.} \end{figure}

As shown in Fig. \ref{fig3}, the number of emitters that de-excited to the lower energy level (emitted radiation) would be maximum at the edge. It may not be hard to infer that this aspect of population dynamics is independent of time, although the actual values of $\langle\hat{n}_{a}(t)\rangle$ increases with time as presented in Fig. \ref{fig2}. Owing to the presumption that the non-radiative emission would be quite negligible near the edge, this result suggests as many as required emitters can participate in the emission process without necessarily decaying to the state other than the one earmarked as the lower. This apprehension then insinuates the prospect of confining intense radiation with frequency very close to edge frequency. With technical possibility of channeling out this radiation by altering the makeup of photonic crystal, it appears promising looking for a means of generating intense light with nonclassical features by means of double-band structure as in correlated emission laser \cite{jmo5433}.

We also seek to explore the nature of photon number correlation that can be utilized to expose quantitative departure from Poisson statistics that designates coherent light \cite{mandle}. This measure of departure is denoted by Mandel's Q-factor:
\begin{align}\label{ve34}M={\langle(\Delta\hat{n})^{2}\rangle-\langle\hat{n}\rangle\over\langle\hat{n}\rangle},\end{align} where $\hat{n}$ is the corresponding photon number operator \cite{ol4205}. It is thus instructive evaluating the likelihood of the manifestation of nonclassical fluctuations in terms of the statistics the radiation reveals, which can be expressed as
\begin{align}M(t)={\langle\hat{n}_{r}^{2}(t)\rangle -\langle\hat{n}_{r}(t)\rangle^{2}-\langle\hat{n}_{r}(t)\rangle\over\langle\hat{n}_{r}(t)\rangle}.\end{align}

To obtain the accompanying Mandel's $Q$-factor, since one needs to determine $\langle\hat{n}^{2}_{r}(t)\rangle$, it is possible to verify applying the method introduced for Gaussian variables \cite{method} in which
\begin{align}\langle\hat{n}_{r}^{2}(t)\rangle=2\langle\hat{a}^{\dagger}(t)\hat{a}(t)\rangle^{2} +\langle\hat{a}^{\dagger}(t)\hat{a}(t)\rangle\end{align}
upon taking into account
\begin{align}\label{incor}\langle\alpha^{2}(t)\rangle=\langle\beta^{2}(t)\rangle=0\end{align}
that
\begin{align}M(t)=\sinh^{2}(G(\omega_{k})t).\end{align}
%

One can then deduce in view of Fig. \ref{fig2} that the radiation confined in PBG exhibits super-Poisonian photon statistics in practically relevant context. It is so reasonable to affirm as the radiation alone may  not possess quantum content at steady state, which somehow supports the outcome reported for ensemble of two-level atoms placed in a lossless cavity \cite{jmo551683, pra241460}. Since Mandel's $Q$-factor turns out to be the same as mean photon number, it is sensible proclaiming the absence of induced coherent superposition among emitted radiations that led to manifestation of chaotic nature. If one wishes to construe nonclassical correlations among emitted radiation from different emitters or at different times, one needs to prepare the emitters in a certain initial coherent superposition or pump the intended scheme externally in the same way as in the cavity scenario \cite{sariv, jmo551587}.

In the same regard, discussion on correlated systems of atom(s) and radiation has been gathering a considerable attention over the years \cite{jmo551683}, where mapping quantum states of the carrier onto the storing device is one of the issues that requires in-depth investigation \cite{prl831319, pra475138}. The coupling among atoms in the ensemble via exchange of spontaneously emitted photons, for instance, is found to be responsible for entanglement and spin squeezing \cite{pra475138, prl794782}.  There is also an evidence that the quantum state of the ensemble of two-level atoms placed in cavity can entangle with spontaneously emitted radiation \cite{jmo551683}. With this in mind, it appears imperative enquiring whether the process of spontaneous emission followed by absorption in the ensemble of two-level emitters embedded onto crystal structure results in correlation capable of forging entanglement?

To gain more insight in this direction, Einstein-Podolsky-Rosen type quadrature operators for emitted radiation and occupancy of the lower energy level, 
\begin{align}\hat{X}_{r}={1\over2}\big(\hat{a}^{\dagger}+\hat{a}\big); ~~~~\hat{X}_{a}={1\over2}\big(\hat{b}^{\dagger}+\hat{b}\big),\end{align}
\begin{align}\hat{P}_{r}={i\over2}\big(\hat{a}^{\dagger}-\hat{a}\big); ~~~~\hat{P}_{a}={i\over2}\big(\hat{b}^{\dagger}-\hat{b}\big),\end{align}
are used \cite{pr47777, pra74043816}. In view that radiation and spin properties are put on equal footing and the accompanying $Q$-function is Gaussian in nature, it is advisable applying the criterion set earlier to quantify continuous variable bipartite entanglement \cite{pra74043816}.

According to the suggested criterion, quantum state of the composite system that can be described by
\begin{align}\hat{V}={1\over\sqrt{2}}\big(\hat{a}^{\dagger}+\hat{a}-\hat{b}^{\dagger}-\hat{b}\big)\end{align} 
is said to be entangled provided that
\begin{align}\Delta V^{2}<1\end{align} 
wherein
\begin{align}&\langle\hat{a}^{\dagger}(t)\hat{a}(t)\rangle +\langle\hat{b}^{\dagger}(t)\hat{b}(t)\rangle +\langle\hat{a}^{\dagger}(t)\rangle\langle\hat{b}^{\dagger}(t)\rangle +\langle\hat{a}(t)\rangle\langle\hat{b}(t)\rangle\notag\\&< \langle\hat{a}^{\dagger}(t)\hat{b}^{\dagger}(t)\rangle +\langle\hat{a}(t)\hat{b}(t)\rangle+\langle\hat{a}^{\dagger}(t)\rangle\langle\hat{a}(t)\rangle
\notag\\&+\langle\hat{b}^{\dagger}(t)\rangle\langle\hat{b}(t)\rangle.\end{align}
It is notably possible to see with the aid of Eq. \eqref{ape4} that
\begin{align}\label{cor1}\langle\hat{a}(t)\rangle=\langle\hat{b}(t)\rangle =\langle\hat{a}^{\dagger}(t)\hat{b}(t)\rangle=0,\end{align}
\begin{align}\label{cor2}\langle\hat{a}(t)\hat{b}(t)\rangle=\sinh(G(\omega)t)\cosh(G(\omega)t).\end{align}
Based on Eqs. \eqref{meanr}, \eqref{meana}, \eqref{incor}, \eqref{cor1}, and \eqref{cor2}, one can then find
\begin{align}\Delta V^{2}(t)& =1+2\sinh(G(\omega_{k})t)\notag\\&\times\big[\sinh(G(\omega_{k})t)-\cosh(G(\omega_{k})t))\big].\end{align}

\begin{figure}[htb]
\centerline{\includegraphics [height=6.5cm,angle=0]{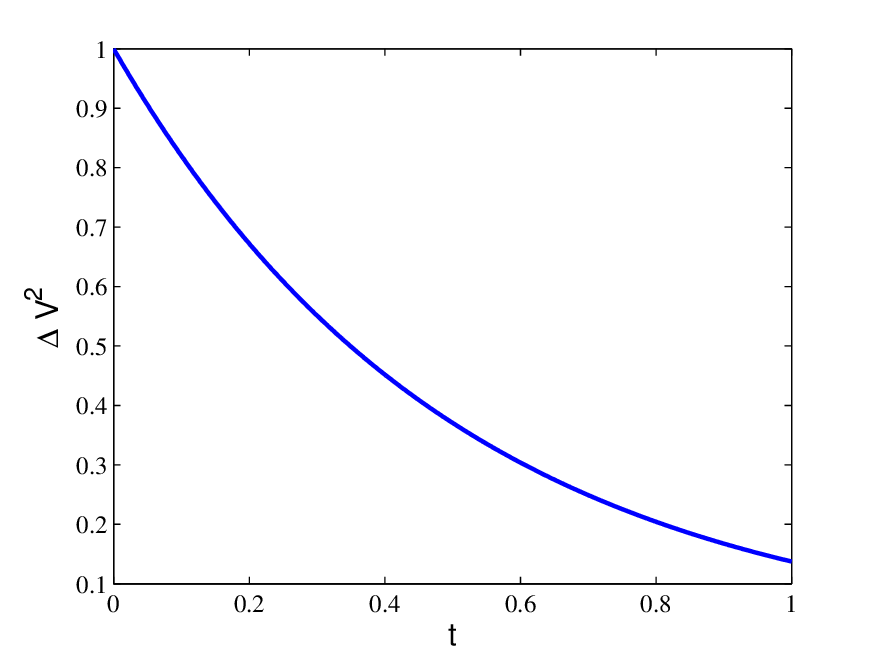}}
\caption {\label{fig5}Variance of the quadrature operators against time for $k = 1.12k_{0}$.} \end{figure}

As clearly seen in Fig. \ref{fig5},  the system under consideration can exhibit strong induced emitter-radiation entanglement whose degree increases with time. This outcome entails that the emission-absorption interplay gets a better chance to be correlated as time of interaction increases, which can also be interpreted as the emitters that participated in the interaction gets enhanced chance to absorb the available photon and re-emit it until the occupancy of the upper and lower energy levels becomes fuzzy. Since the intensity of confined radiation increases with the time as shown in Fig. \ref{fig2}, the heating of the overall setup can lead to thermal fluctuations that have a capacity to affect the physical makeup of the system and so requires further in-depth consideration.

\begin{figure}[htb]
\centerline{\includegraphics [height=6.5cm,angle=0]{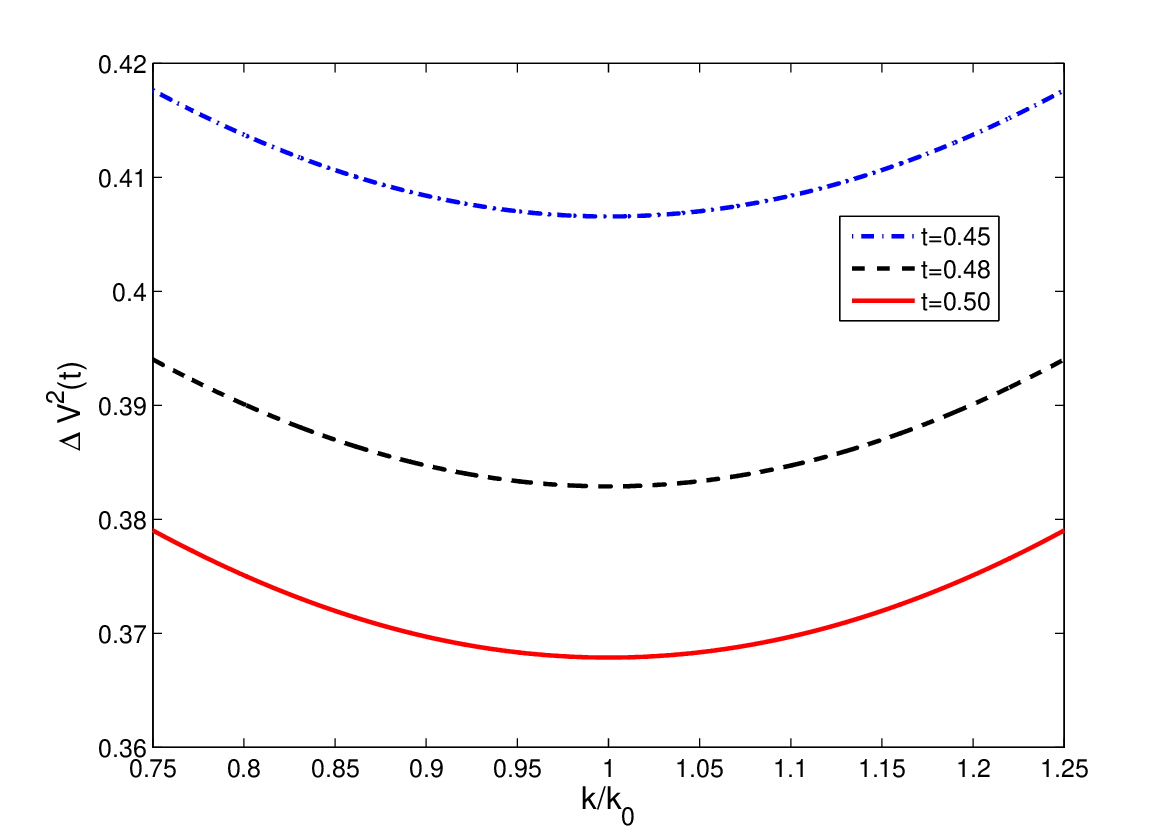}}
\caption {\label{fig6}Variance of  EPR type quadrature operators against $k/k_{0}$ at different times.} \end{figure}

It is also possible to see from Fig. \ref{fig6} that the degree of entanglement increases with approaching the edge. Owing to the observation that the non-radiative spontaneous emission is one of the phenomena by which the nonclassical features would be lost, it is comprehendible to realize that the degree of entanglement is maximum when non-radiative damping is minimum. The illustrated characteristics of entanglement can also be linked to the strength of spontaneous emission near the edge that has a potential to expedite successive correlated emission-absorption events responsible for entanglement. Further scrutiny in this direction reveals that such behavior of induced entanglement is independent of time, although the actual degree of entanglement can be different at different times. It might also be possible to conjecture that the degree of entanglement can be controlled by adjusting the places the emitters are placed, which could be compelling in practical context.

It may worth noting that the inherent nonclassicality can also be captured by intensity correlation. To ascertain this proposal, it turns out to be appealing to construct various coupled operators such as \cite{books, prl96050503} 
\begin{align}\label{qpl148}\hat{\ell}_{+}=\hat{a}^{\dagger}\hat{b}^{\dagger}+\hat{a}\hat{b};\hspace{.7cm} \hat{\ell}_{-}=i(\hat{a}^{\dagger}\hat{b}^{\dagger}-\hat{a}\hat{b})\end{align}
in which the combined state is taken as entangled in case \cite{prl96050503}
\begin{align}\label{qpl149}\big|\langle\hat{a}\hat{b}\rangle\big| >\sqrt{\langle\hat{a}^{\dagger}\hat{a}\rangle\langle\hat{b}^{\dagger}\hat{b}\rangle},\end{align}
which can also be rewritten as
\begin{align}\label{qpl150}\big|\langle\hat{a}\hat{b}\rangle\big| >\sqrt{\langle\hat{n}_{a}\rangle\langle\hat{n}_{r}\rangle}.\end{align}

Upon defining afterwards
\begin{align}\label{eh}E_{c}={\sqrt{\langle\hat{n}_{a}\rangle\langle\hat{n}_{r}\rangle}\over C_{c}},\end{align}
where
\begin{align}\label{pnc1}C_{c}=\big\langle\hat{a}(t)\hat{b}(t)\big\rangle=\sinh(G(\omega)t)\cosh(G(\omega)t)\end{align}
quantification of entanglement is epitomized by
\begin{align}\label{eh2}E_{c}<1.\end{align}

\begin{figure}[htb]
\centerline{\includegraphics [height=6.5cm,angle=0]{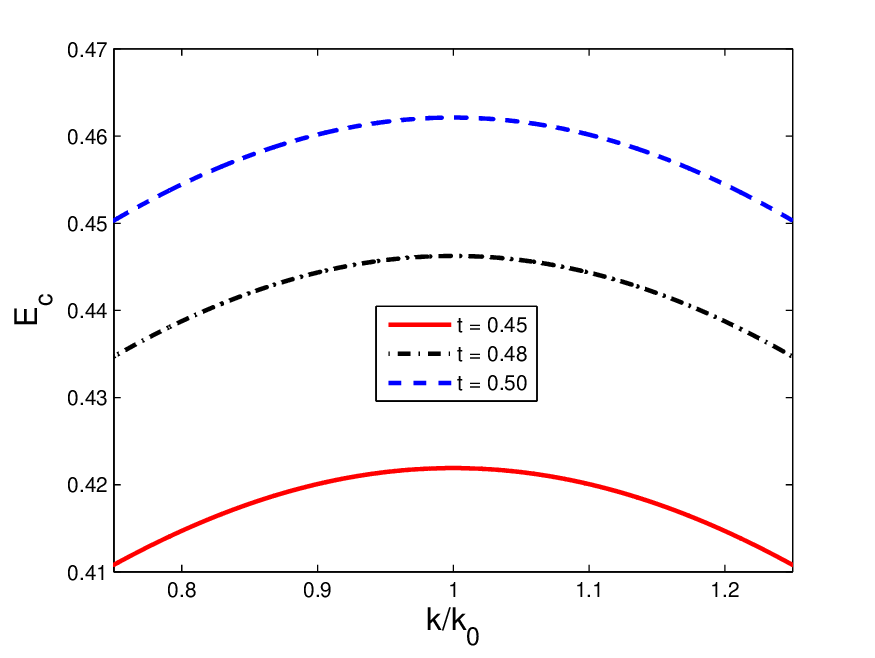}}
\caption {\label{fig7}Nonclassical photon number correlation against $k/k_{0}$ at different times.} \end{figure}

One can deduce from Eq. \eqref{eh} or Fig. \ref{fig7} that $E_{c}$ would be closer to zero in case coherent correlation is very high when compared to mean photon numbers. Similar verdict can also be reached upon when at least one of the photon numbers is found to be close to zero irrespective of the value of correlation. It is then required to critically compare photon number with correlation prior to arriving at conclusion \cite{jpb42215506}. With this in mind and without going into details of its strength, one can see from Fig. \ref{fig7} that there can be strong nonclassical correlation between quantum states of emitted radiation and occupancy of the lower energy level. This observation supports the claim that the emerging state of the interaction of ensemble of two-level atoms embedded into PBG with confined radiation can be significantly entangled.


In conclusion, interaction of an ensemble of excited two-level emitters embedded onto arrays of dielectric structure with the emitted radiation is found to lead to robust entanglement, where the degree of entanglement, photon statistics, and population dynamics depend on interaction time and deviation of the frequency of the emitted radiation from band edge frequency. The degree of entanglement and intensity of the radiation is shown to increase with interaction time that can be linked to the number of emitters involved in the interaction and prospect of witnessing strongly correlated emission-absorption events over time. Particularly, the degree of entanglement is found to be significantly enhanced when compared to the corresponding cavity scenario \cite{jmo551683}. Even though confining electromagnetic field leads to nonclassical coupling, the emitted radiation alone is shown to exhibit super-Poissonian and chaotic photon character at relevant time regime. 
Besides, intensity of the generated electromagnetic field is found to be significant where and when the degree of entanglement is large, which can be taken as a positive indication for further scrutiny and potential application of the system under consideration. With recent development that makes possible the radiation to be channeled out by altering the makeup of the crystal, it is envisaged as embedding emitters onto regular structure of various types can be taken as a fertile ground for analyzing the emerging nonclassical correlations between emitters and radiation. Despite the involved challenges, this study highlights the prospect of demonstrating robust entanglement of combined state of custom designed emitters and radiation confined in PBG that can be applicable in the area of quantum electrodynamics and quantum information processing.

\end{document}